# Electronic Band Structures and Excitonic Properties of Delafossites: A *GW*-BSE study

Xiaoming Wang*, Weiwei Meng and Yanfa Yan*
*Department of Physics and Astronomy, and Wright Center for Photovoltaic Innovation and Commercialization, The University of Toledo, Toledo, OH 43606, United States*
e-mail address: wxiaom86@gmail.com, yanfa.yan@utoledo.edu

We report the band structures and excitonic properties of delafossites $CuMO_2$ (M = Al, Ga, In, Sc, Y, Cr) calculated using the state-of-the-art *GW*-BSE approach. We find that all the delafossites are indirect band gap semiconductors with large exciton binding energies, varying from 0.24 to 0.44 eV. The lowest and strongest exciton, mainly contributed from either Cu $3d \rightarrow$ Cu $3p$ (Al, Ga, In) or Cu $3d \rightarrow$ M $3d$ (M = Sc, Y, Cr) transitions, is always located at L point of the rhombohedral Brillouin zone. Taking the electron-hole effect into account, our theoretical band gaps exhibit nice agreement with experiments.

## I. INTRODUCTION

Nowadays transparent conductive oxides (TCO) are widely used in displays, solar cells and light emitting diodes. Such materials require optical transparency with band gap over ~3 eV and good *p*- and *n*-type dopability. Most of the high-performance TCOs are *n*-type semiconductors, such as $In_2O_3$, ZnO and $SnO_2$. Their use is rather restricted due to lack of the *p*-type counterparts. Realization of *p*-type TCOs is challenging but quite important since many semiconductor devices are based on *p-n* junctions. *p*-type conductivity in combination of transparency was achieved in 1997[1] in thin films of $CuAlO_2$, which crystallizes in delafossite structure belonging to *R3m* space group. Subsequently, *p*-type TCOs has been discovered in other delafossites ($CuMO_2$, M = Ga, In, Sc, Y, Cr).[2–6]

There are many research reports on the electrical and optical properties of delafossites over the past two decades, both theoretical and experimental. However, the agreement with each other is rather limited. The discrepancy comes from the sample fabrication details, measurement conditions and also the different levels of theories used. Among the delafossites, $CuAlO_2$ is by far the most studied, which is found to be an indirect band gap semiconductor. Early measurements demonstrated an indirect band gap of 1.65-2.1 eV[7–10], which is argued and classified as defect levels.[11–14] Recent experiments[12,13,15] show the indirect band gap to be ~3 eV, while the direct optical transition at ~3.5 eV is generally accepted. On theoretical side, two important aspects in delafossites require careful attention when comparing the theoretical fundamental band gaps with experiments. One is that the direct transition at Γ point is optically forbidden due to parity reason[16]. The direct optical gap is found at L point, though the band gap at Γ may be even smaller, for example in $CuInO_2$. The other one is the large exciton binding energy found in these materials[17–20]. The delafossite structure can be considered as layered structure with the metal planes connected by O-Cu-O bonds. Electrons are highly confined in the planes resulting in weak screening in the cross-plane direction and enhancing the electron-hole interactions. Note, that the difference between the theoretical fundamental gap and the optical gap is the exciton binding energy.

Density functional theory (DFT) calculations[16,21] based on semilocal exchange-correlation functionals, namely LDA or GGA, predict the indirect and direct band gaps of $CuAlO_2$ to be ~2.0 eV and ~2.7 eV, respectively. The underestimation is a well-known deficiency of DFT as a result of the self-interaction error. Various approaches beyond DFT have been developed to improve the description of the band gaps. The simplest one, both effective and efficient, is the DFT+U scheme[22,23], where an on-site Hubbard U term is added to the localized orbitals (Cu $3d$ in $CuAlO_2$) to correct the electron correlations. The DFT+U method effectively opens the band gap of $CuAlO_2$, the size of which, however, greatly depends on the U parameter.[17,24,25] This makes the method less favorable in the sense of prediction. An alternative is to use hybrid functionals, which incorporate a portion of Hartree-Fock (HF) exact exchange with exchange and correlation from semilocal DFT functionals. Hybrid functionals, balancing out the overestimation of HF and the underestimation of DFT, are found to yield good results for simple semiconductors.[26–30] However, for more complex systems or semiconductors with *d* electrons, the general validity of hybrid functional is not clear. As for $CuAlO_2$, the band gap at L, which corresponds to the direct optical transition, was predicted to be 4.5 eV[11] and 4.8 eV[31] from B3LYP[32] and PBE0[33,34] hybrid functionals. Even including the exciton binding energy of ~0.5 eV[17], these numbers are still too large. sX-LDA[28] underestimates the value to be 2.95 eV.[35] HSE06[36] functional seems to predict reasonable band gap of ~4.1 eV[25,31], which is comparable to experimental value considering the excitonic effect. However, the HSE06 band gap of $CuScO_2$ is 3.72 eV as in our calculations, significant underestimation of the experiment values of 4.24-4.35 eV[18,19].

A more involved theoretical method is many-body perturbation theory with *GW* approximation[37,38]. The *GW* method, including the many-body electron-electron interaction, in general predicts more accurate band structures of solids.[39–43] In practice, single shot *GW* or $G_0W_0$ as a perturbative quasiparticle correction to LDA band structure is widely used and shown to yield accurate band gaps for some typical semiconductors,[38,43] but generally still gives an underestimation and the error can be large for systems with shallow *d* states[43]. Different levels of *GW* approximation exist to improve the agreement with experiments. Partial self-consistency on the Green's function *G* only ($GW_0$) and on both *G* and screened Coulomb interaction *W* (*GW*) can further open the band gap.[43] The $GW_0$ scheme, with the screening properties fixed at random phase approximation (RPA) level, shows excellent agreement with experiments[43,44], stemming from the fact that the dielectric properties of many systems can be predicted quite well at RPA[45]. Further update of *W*, as in the *GW* scheme, is found to give too large band gaps of almost all materials due to underestimation of the screening.[43] Another approach solving the deficiency of $G_0W_0$ is based on the generalized Kohn-Sham (gKS) scheme, which uses hybrid functional wave functions as a starting point.[42] The gKS scheme is also a practical solution of *GW* calculations if LDA or GGA predict completely unreasonable wave functions, as we will see it is the case for $CuCrO_2$. All the above *GW* approximations depends on the starting point of mean field wave functions, different starting points yield varied results. Quasiparticle self-consistent *GW* (QS*GW*)[41], developed to optimized the starting point self-consistently, shows systematically improved band structures of many materials, but suffers from an overestimation of the fundamental gaps. The overestimation can be partly remedied by including the vertex correction in the dielectric matrix.[46] $G_0W_0$@LDA calculations[31,47] predict the indirect and direct band gaps of $CuAlO_2$ of 3.1 eV and 3.4 eV, respectively, which are obviously underestimated given the fact of large exciton binding energy. By using wave functions from scCOHSEX as a starting point, the $G_0W_0$ band gaps are changed to 5.0 eV and 5.1 eV[31,47]. A similar large band gap was also obtained by QS*GW* calculations.[48] Contrary to all other theories, $G_0W_0$@scCOHSEX predicted the conduction band minimum (CBM) located at L instead of Γ. The large discrepancy of the band gap compared with experiments was claimed attributed to large polaronic effect which shrinks the band gap by 1.2 eV according to a simple model calculation. By adding up both the excitonic and polaronic contributions, agreement with experiments can be eventually obtained. However, we note that the phonon frequency of $CuAlO_2$ is $\omega_{LO}$ = ~80-90 meV, which is responsible for the polaron shift[49]. It is obvious that the phonon modes in the relevant energy is practically not active since the measurements are conducted at room temperature[13] or even as low as 20 K[12]. Therefore, the large band gap must be ascribed to the underestimation of the dielectric constant in scCOHSEX calculation, as found similar to QS*GW*.[41]

As we see there is yet no consistent approach appropriately describing the band structures of delafossites. In addition, the measured exciton binding energies of delafossites are as large as 0.3-0.5 eV[18–20]. However, there is only one report calculating the exciton absorption of $CuAlO_2$ by solving the Bethe-Salpeter equation[50–53] (BSE) based on LDA+U band structures.[17] Though tuning the U parameter can reproduce the experimental band structures and the following BSE could give a reasonable estimate of the exciton binding energy, we claim that more accurate and parameter free calculations are desired from the first-principles point of view. Here, we present $GW_0$ calculations on the band structures of ($CuMO_2$, M=Al, Ga, In, Sc, Y) with PBE[54] wave functions as the starting point. For $CuCrO_2$, PBE predicts completely wrong orders of Cr 3*d* bands, hence, HSE is adopted. The BSE is solved on top of the $GW_0$ quasiparticle band structures. The exciton binding energies are found to vary in range of 0.24-0.44 eV for different delafossites. Our *GW*-BSE results agree very well with experiments.

## II. METHODOLOGY

All the DFT, *GW* and BSE calculations are performed using the VASP code[40,55,56] and projector augmented-wave (PAW)[57] potentials. For Ga and In, the potentials with semicore 3*d* and 4*d* electrons placed in valence are used, while for Sc, Y and Cr, we use 3*s*3*p*, 4*s*4*p* and 3*s*3*p* shells in valence. We use Perdew, Burke, Ernzerhof (PBE)[54] exchange-correlation functional to relax the atomic positions while experimental lattice parameters are used for all the delafossites. The force tolerance is 0.01 eV/Å. A k-grid of 6×6×6 is found to converge the band structures within 0.01 eV both for DFT and *GW* calculations, and a kinetic energy cutoff of 500 eV is used to expanding the wave functions.

We carefully examine the convergence of the *GW* calculations, especially the energy cutoff of the response functions $E_\varepsilon$ and the number of empty states *N*, since we have *d* bands at the band edges of delafossites. It is demonstrated that large $E_\varepsilon$ and *N* are needed to get reliable results of localized states.[58] Usually, the band gap converges much faster than the exact band energies due to the error cancellation. Figs. 1a and b show the band energy and band gap convergence with respect to $E_\varepsilon$ and *N*, respectively, at Γ for $CuAlO_2$. It is obvious that $N = 100$ and $E_\varepsilon = 200$ eV are enough to converge the band gap within 0.02 eV. The same accuracy for the exact band energy can only be obtained for at least $N = 1500$ and $E_\varepsilon = 400$ eV. Contrary to the behavior of Γ, the convergence is more difficult for L, where the CBM converges much faster than VBM, as shown in Figs. 1c and d. As a result, the band gap convergence slows down. We attribute this to the different band characters of VBM and CBM. The VBM is mostly contributed by Cu 3d orbitals at all k points, though a hybridization with O 2p orbitals exists, as shown in Fig. 3a. The CBM is dominated by Cu 3d orbitals at Γ but Cu 3p orbitals at L. Since 3d orbital is more localized than 3p orbital, the CBM at L needs more effort to converge. In a word, we choose $N = 1000$ and $E_\varepsilon = 300$ eV in our GW calculations. This restricts the band gap error within 20 meV. The dielectric function is evaluated at 100 frequency grids on real axis. The self-energy is calculated using the full frequency integration method.[40] After obtaining the GW eigenvalues on a regular k mesh, the band structures on high symmetry lines are interpolated using WANNEIR90 code[59].

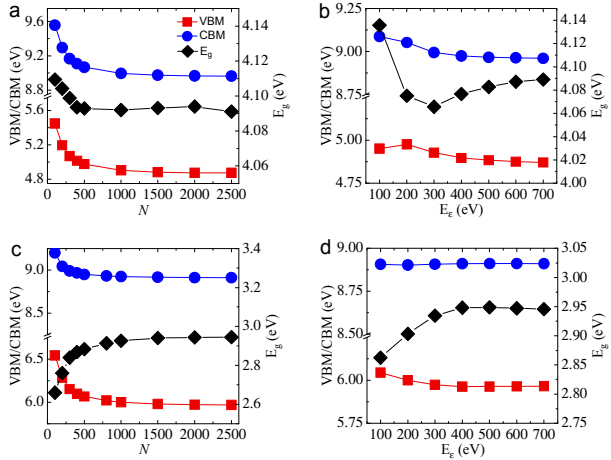

FIG. 1 Convergence of the band energy and band gap of $CuAlO_2$ with respect to the number of empty states $N$ and energy cutoff of the response function $E_\varepsilon$ at (a), (b) Γ point and (c), (d) L point. The calculations are conducted at a k grid of 2×2×2, since the convergence behavior is insensitive to the k mesh. For (a) and (c), $E_\varepsilon = 700$ eV, while for (b) and (d), $N = 2500$.

The excitonic properties are determined by solving the BSE with Tamm-Dancoff approximation[60] on top of the GW quasiparticle band structure. The indirect transitions are not included in the present study. Two occupied and one unoccupied bands are used and found enough to converge the lowest electron-hole excitation energy $\Omega_1$ which determines the exciton binding energy. It is usually required much finer k mesh in BSE calculations to obtain accurate exciton spectrum, since excitons are very localized in reciprocal space. However, the computation demand is prohibitive even for the GW calculations for many k points. To this end, we use a model BSE (mBSE) scheme[61]. The dielectric function was fitted to GW calculations with the form:

$$\varepsilon^{-1}(\boldsymbol{q}) = 1 - (1 - \varepsilon_\infty^{-1})e^{\frac{-\pi^2|q|^2}{q_{TF}^2}} \quad (1)$$

Where $\varepsilon_\infty$ is the ion-clamped static dielectric constant which is evaluated on RPA level. $\varepsilon_\infty$ for different delafossites are listed in Table I. $q_{TF}$ is the Thomas-Fermi screening length, which is found nearly constant of 1.1 for all the delafossites by fitting. The quasiparticle eigenvalues are evaluated on a finer k mesh through wannier interpolation. We argue that wannier interpolation is more accurate than scissor shift technique[61] and essential for non-uniform corrections of the GW bands. The PBE wave functions at finer k mesh are used. However, for $CuCrO_2$, PBE wave function is unreliable. The PBE+U method are adopted. We find that the exciton spectrum is insensitive to the U parameter, since the eigenvalues are fixed at GW level. Fig. 2a compares the imaginary part of the dielectric function $\varepsilon_2$ (Hereafter, we ignore the polarization dependence and use $\varepsilon = (\varepsilon_{xx} + \varepsilon_{yy} + \varepsilon_{zz})/3$ ) of $CuAlO_2$ calculated by GW-BSE and the mBSE method using 6×6×6 k mesh. As can be seen, mBSE reproduce the GW-BSE curve quite well. A finer k mesh of 12×12×12 is used in all the mBSE calculations to determine the exciton binding energy. The estimated error is 0.01 eV, as show in Fig. 2b.

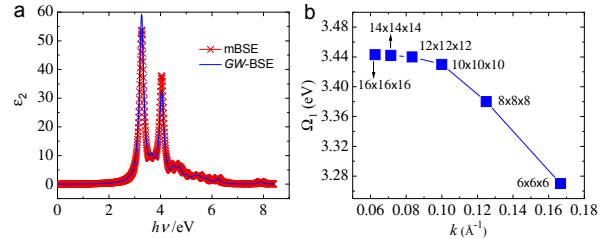

FIG. 2 (a) The comparison of the imaginary dielectric function $\varepsilon_2$ calculated by GW-BSE and mBSE. (b) The convergence of the lowest exciton energy $\Omega_1$ with respect to the k mesh using mBSE method.

## III. RESULTS AND DISCUSSION

The band gaps and exciton binding energies of delafossites calculated by the present GW-BSE approach is summarized in Table I, PBE band gaps and experimental direct optical gaps are also included for comparison. Only $CuScO_2$ is predicted to be direct band gap semiconductor by PBE. Our GW calculations show that all the delafossites are indirect band gap semiconductors. We also include the lowest excitation energy $\Omega_1$ in Table I, which, in general, agrees quite well with the experimental optical gaps. The exciton binding energy $E_{xb} = E_g^d - \Omega_1$ shows variation for different delafossites with largest of 0.44 eV for $CuAlO_2$ and smallest of 0.24 for $CuScO_2$. These data agree with the reported BSE@LDAU calculation of 0.47 eV[17] for $CuAlO_2$ and the experimental data of 0.3-0.38 eV[18,19] for $CuScO_2$. In what follows, we present more details about the band structures and excitonic properties of different delafossites.

TABLE I. Band gaps of delafossites. CMO is shot for $CuMO_2$, where M is the first letter of Al, Ga, In, Sc, Y and Cr, respectively. $E_g^d$ is direct band gap, $E_g^i$ is indirect band gap, $\Omega_1$ is the lowest electron-hole excitation energy, $E_{xb}$ is exciton binding energy, and $E_o$ is the direct optical gap from measurements. $\varepsilon_\infty^{RPA}$ is the static ion-clamped dielectric constant evaluated on RPA level with local field effect included.

|  |  | CAO | CGO | CIO | CSO | CYO | CCO |
|---|---|---|---|---|---|---|---|
| PBE | $E_g^d$ | 2.68 | 2.58 | 3.09 | 2.21 | 2.69 | 1.30 |
|  | $E_g^i$ | 1.92 | 0.92 | 0.40 | --- | 2.67 | 0.96 |
| GW | $E_g^d$ | 3.88 | 3.80 | 4.49 | 4.22 | 4.49 | 3.42 |
|  | $E_g^i$ | 3.55 | 2.48 | 1.89 | 4.16 | 4.35 | 3.29 |
|  | $\Omega_1$ | 3.44 | 3.47 | 4.07 | 3.98 | 4.16 | 3.11 |
|  | $E_{xb}$ | 0.44 | 0.33 | 0.42 | 0.24 | 0.33 | 0.31 |
|  | $\varepsilon_\infty^{RPA}$ | 5.3 | 5.6 | 5.0 | 5.5 | 4.6 | 4.9 |
| EXP[a] | $E_o$ | 3.46 | 3.53 | 3.90 | 3.93 | 3.7 | 3.08 |
|  |  | 3.47 | 3.56 | 4.15 | 3.94 | 3.8 | 3.14 |
|  |  | 3.53 | 3.60 | 4.45 |  | 3.5 | 3.18 |

[a]Experimental data are from Refs[2,3,5,6,12,13,15,18,19,62–69].

## A. $CuAlO_2$, $CuGaO_2$ and $CuInO_2$

As mentioned in the introduction, different levels of self-consistency in GW calculations can yield quite different band structures. We confirmed this by examining the direct band gap $E_g^d$ of $CuAlO_2$ at L, as shown in Table II. For all the GW calculations, we find the CBM is located at Γ, with 0.2-0.3 eV lower than that at L. Our $G_0W_0$ band gap of 3.5 eV agrees with 3.4 eV reported by Trani et al[31]. Though the present QSGW calculations are different from what they called scGW ($G_0W_0$@scCOHSEX) in Refs[31,47], the obtained band gaps of 5.1 eV are the same. This is because the wave functions produced by the two different approaches are extremely close.[70] Christensen et al[48] reported a smaller band gap of 4.55 eV with the same QSGW method. We attribute the difference to the different basis sets used in the GW calculations. Plane waves and PAW potentials are used in our calculations while they adopt full-potential linear muffin-tin orbitals (LMTO). As can be seen from Table II, $E_g^d$ is increased in the order of $G_0W_0$ < $GW_0$ < $GW$ < QSGW. The self-consistency of the screened Coulomb potential W reduces the electronic screening, as indicated by the decreasing of the dielectric constant, hence, deteriorating the agreement with experiments.[43] The RPA dielectric constant is 5.3, close to the experimental value of 5.1[71]. Hence, a reasonable evaluation of the quasiparticle band structure should be expected to be obtained at $W_0$ level.

TABLE II. The calculated GW band gaps $E_g^d$ of $CuAlO_2$ at L with different levels of self-consistency. $\varepsilon_\infty^{RPA}$ is the static ion-clamped dielectric constant evaluated on RPA level with local field effect included.

|  | $G_0W_0$ | $GW_0$ | $GW$ | QSGW |
|---|---|---|---|---|
| Present work | 3.5 | 3.9 | 4.3 | 5.1 |
| Refs. | 3.4[31] |  |  | 5.1[31,47], 4.55[48] |
| $\varepsilon_\infty^{RPA}$ | 5.3 | 5.3 | 4.1 | 3.4 |

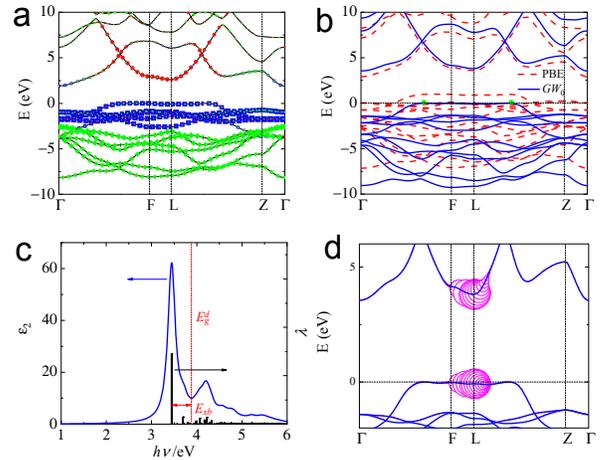

FIG. 3 Band structure of $CuAlO_2$. (a) PBE fat bands with projections on Cu 3d (blue squares), Cu 3p (red circles) and O 2p (green diamonds) orbitals, respectively. (b) Comparison of the PBE (red dash) and $GW_0$ (blue solid) band structures. The green dots are the VBM positions along ΓF and LZ lines. (c) The imaginary dielectric function $\varepsilon_2$ and oscillator strength $\lambda$ as a function of photon energy $h\nu$. (d) The lowest exciton amplitude (magenta circle) plotted as a fat-bands style.

Fig. 3a displays the fat bands of $CuAlO_2$ with projections on Cu 3d, 3p and O 2p orbitals. The Cu 4s orbitals have a small contribution to the CBM at Γ, which is not shown here. It is clear that the VBM for all the k path is mainly contributed from the Cu 3d orbitals, with part hybridization with O 2p states. On the contrary, the CBM exhibits distinct band characters with Cu 3d and 3p predominates for Γ and L, respectively. Fig. 3b compares the $GW_0$ band structure with that of PBE. The VBM is located along the ΓF line, as indicated by the green dot, which we denote as A site. Hence, the indirect band gap is measured from A to Γ. The VBM along LZ line, denoted as B, is only 5 meV lower than that of A. The $GW_0$ and PBE band structures are aligned with respect to the averaged electrostatic potential, since the charge density is not changing during $GW_0$ calculations. As a general trend, the $GW_0$ corrections shift the conduction and valence bands up and down, respectively. The band renormalizations, however, varies at different k points. The VBM correction $\Delta E_{VBM}$ is 1.0 eV at L while 1.5 eV at Γ. The $\Delta E_{CBM}$ is much smaller with 0.65 eV at Γ and 0.2 eV at L. The non-uniform renormalization of the bands invalidates the scissor shift technique in the delafossite structures.

Based on the $GW_0$ band structure, the exciton properties are evaluated by solving the BSE. The exciton oscillator strength and the imaginary part of the dielectric function $\varepsilon_2$ are plotted in Fig. 3c. The first peak of $\varepsilon_2$, corresponding to the lowest excitation energy of 3.44 eV, which is smaller than the related band gap $E_g^d$ of 3.88 eV, indicates a bound exciton with a binding energy of 0.44 eV. Such high exciton binding energy is attributed to the localization of exciton wave functions in real space[17] and the weak screening environment. The exciton wave functions can be expressed in an coupled electron-hole pair configurations: $|S\rangle = \sum_{vck} A_{vck}^S |vc\rangle$, where $v$ and $c$ are for valence and conduction states, $A_{vck}^S$ is the exciton amplitude with the corresponding excitation energy $\Omega_S$. We plot $|A_{vck}^1|$ for the lowest excitation as fat bands[72], as shown in Fig. 3d. The circle size denotes the absolute value of the amplitude. In practice, BSE is solved on a regular $k$ mesh. To calculate $A_{vck}^S$ for an arbitrary $k$ point as for the purpose of plot, we adopt the Gaussian mapping:

$$A_{vck}^S = \sum_{k'} \frac{A_{vck'}^S w_{k'}}{(\sigma\sqrt{2\pi})^3} \exp\left(-\frac{|k-k'|^2}{2\sigma^2}\right) \quad (2)$$

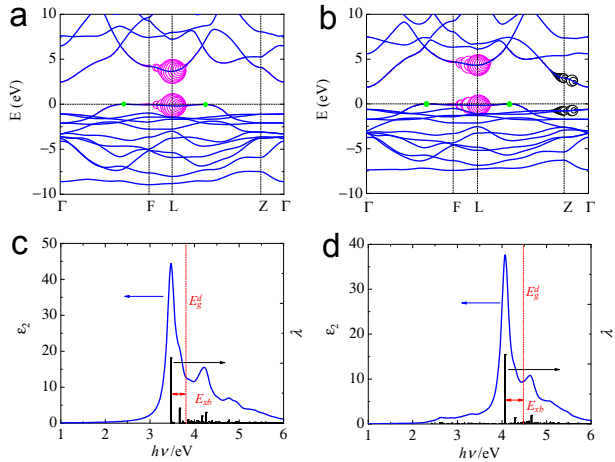

FIG. 4 (a), (b) $GW_0$ band structures with fat-bands feature of excitons and (c), (d) dielectric function $\varepsilon_2$ and oscillator strength $\lambda$ of (a), (c) $CuGaO_2$ and (b), (d) $CuInO_2$, respectively. The magenta and black circles in (b) are for excitons with largest oscillator strength and lowest excitation energy, respectively.

Where $k'$ is the k point in irreducible Brillouin zone, $w_{k'}$ is the corresponding weight, $\sigma = 0.05$ is adopted. It is clear from Fig. 3d that the lowest exciton is composed almost exclusively of the transitions between the topmost valence and lowest conduction states around L. Those states are localized Cu 3$d$ and 3$p$ orbitals, as shown in Fig. 3a. Hence, the largest probability of excitations is found to be the Cu 3$d \rightarrow$ Cu 3$p$ transitions on the same Cu atom.[17] The delafossites can be considered as layered structures with excitations confined in the Cu planes. The cross-plane dielectric constant $\varepsilon_\perp$ is calculated to be 4.5, while the in-plane $\varepsilon_\parallel$ is 5.7. The weak electronic screening in the cross-plane direction further enhances the binding the of excitons. After taking account of the exciton binding energy of 0.44 eV, the $GW$-BSE optical band gap is found to be 3.44 eV, which is in good agreement with the experimental value of ~3.5 eV.

The $GW_0$ band structures and exciton properties of $CuGaO_2$ and $CuInO_2$ are shown in Fig. 4. The VBM of both of them are located at A site along the ΓF line, with B site 15 meV and 2 meV lower in energy, respectively, for $CuGaO_2$ and $CuInO_2$. The CBM is located at Γ. The lowest excitation of $CuGaO_2$, corresponding to the first peak of $\varepsilon_2$ in Fig. 4c, is contributed from the transitions around L point, as shown in Fig. 4a. The dominant excitation is still Cu 3$d \rightarrow$ Cu 3$p$ transition similar to $CuAlO_2$. However, the CBM at L also has some Ga 4$s$ character. Hence, Cu 3$d \rightarrow$ Ga 4$s$ transition can also contribute to the lowest excitations. The calculated exciton binding energy is 0.33 eV, which is smaller than that of $CuAlO_2$. The smaller binding energy is related to the spread the exciton wave functions due to Cu 3$d \rightarrow$ Ga 4$s$ excitations and larger dielectric constant, which is calculated as 5.6. For $CuInO_2$, the fundamental direct band gap of 2.5 eV is found at Γ point. Nonetheless, the optical transitions at Γ is dipole-forbidden.[16] The measured optical gap is composed of the transitions at L point, see the magenta circles in Fig. 4b, which is corresponding to the strongest peak of $\varepsilon_2$ in Fig. 4d. However, below this excitation, there are many weak excitations with much smaller oscillator strength. We plot the amplitude of one of these weak excitons in Fig. 4b, as indicated by the black circles. Note, that the amplitude is ten times exaggerated to make suitable to the eye. The weak excitons are composed of the transitions around Z point, contributed from Cu 3$d \rightarrow$ Ga 4$s$ excitations. Since the amplitudes are smaller and the dipole transition probability at Z is also smaller[16], the optical absorption due to these weak excitons can hardly be observed in experiments.

### B. $CuScO_2$ and $CuYO_2$

Unlike the group IIIA (Al, Ga, In) delafossites, the few low-lying conduction states of the IIIB (Sc, Y) ones are contributed by the partially filled $d$ orbitals, as indicated by the red circles in Fig. 5a. The fat-bands features of the valence bands for $CuScO_2$ and $CuYO_2$ are similar to that of $CuAlO_2$ and not shown here. The CBM of both $CuScO_2$ and $CuYO_2$ is located at L, with 0.1 eV lower than that at Γ where the Cu 3$d$ orbitals still predominate. The VBM of $CuScO_2$ is at B site, with 2 meV higher than that of A site. Contrary to experiments[19], $CuScO_2$ is predicted to be an indirect semiconductor with indirect band gap of 4.16 eV from B to L. The direct band gap at L is 4.22 eV, only 0.06 eV larger than the indirect band gap. Hence, such small difference can hardly be detected in UV-vis measurements. The VBM of $CuYO_2$ is located at A site with 35 meV higher than that of B site. $CuYO_2$ is also found an indirect semiconductor with indirect band gap of 4.35 eV from A to L. The direct band gap at L is 4.49 eV.

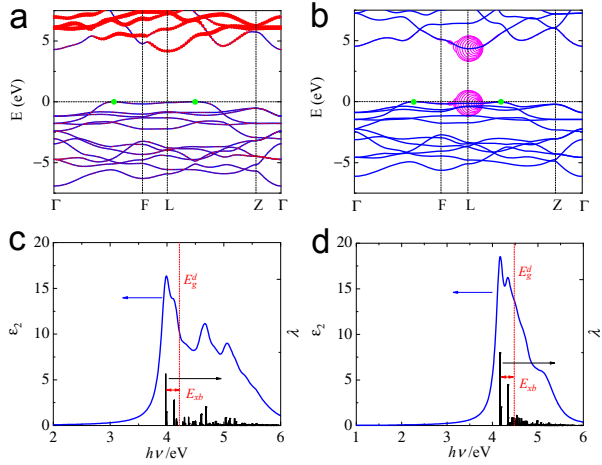

FIG. 5 (a), (b) $GW_0$ band structures and (c), (d) dielectric function $\varepsilon_2$ and oscillator strength $\lambda$ of (a), (c) $CuScO_2$ and (b), (d) $CuYO_2$, respectively. The red circles in (a) denote the fat bands of Sc $d$ orbitals, the magenta circles in (b) show the lowest excitons.

The exciton properties of $CuScO_2$ and $CuYO_2$ are quite similar. The lowest excitations or the first absorption peak in $\varepsilon_2$ as shown in Figs. 5c and d, are composed of the transitions around L mainly from Cu 3$d$ → Sc 3$d$ or Cu 3$d$ → Sc 3$d$ excitations, as in Fig. 5b. A small contribution can be found for Cu 3$d$ → Cu 3$p$ excitations. Since the excitons are dominated by the inter-atom excitations, the binding energy is expected to be smaller than that of the Cu 3$d$ → Cu 3$p$ excitations. The exciton binding energy of $CuScO_2$ and $CuYO_2$ are 0.24 and 0.33 eV, respectively. The smaller exciton binding energy of $CuScO_2$ can be further ascribed to the larger dielectric constant of 5.5 compared to 4.6 of $CuYO_2$. Our calculated exciton binding energy of 0.24 eV and the optical band gap of 3.98 eV agree quite well with the experimental value of 0.3 eV and 3.94 eV[19], respectively. However, extra resonate exciton peaks above the band gap, as shown in Fig. 5c, are not observed in experiments[18,19]. We attribute this discrepancy to the insufficient k point sampling in BSE calculations. Usually, one needs extremely fine k mesh to converge the absorption curve for higher energies. We only check convergence with respect to the k mesh about the first absorption peak, as our focus is on the exciton binding energy. Hence the resonate exciton peaks may not be converged.

### C. $CuCrO_2$

$CuCrO_2$ is different from other delafossites, since electrons are strongly correlated due to the half-filled 3$d$ states of Cr. The ground state was found antiferromagnetic[73] with very small difference in energy to that of ferromagnetic state.[35,74] Here, our calculations are performed on the ferromagnetic state due to its smaller unit cell. The band structure of $CuCrO_2$ is under debate in literatures.[20,69,74–78] The main concern is the Cr 3$d$ positon in the valence bands. PBE calculations predict that both the CBM and VBM are mainly composed of Cr 3$d$ states, as shown in Fig. 6a. Arnold *et al*[75] reported the band structure of $CuCrO_2$ by using X ray spectroscopy combined with DFT+U calculations and showed that the VBM derived mainly from the Cu 3$d$ states. Yokobori *et al*[77] found that Cr 3$d$ states dominate the VBM according to their photoemission spectroscopy (PES) and x-ray absorption spectroscopy (XAS) measurements and also DFT+U calculations. Although the conclusions of the two are contrary to each other, similar valence band spectrum are obtained. The mystery is indeed due to the different U parameters used in their first-principles calculations, since both of their conclusions depend on the DFT+U calculations. In Ref. 74, U of 5 eV and 4 eV were used for Cu and Cr 3$d$ states, respectively, while U of 2 eV was used for both of the atoms in Ref. 76. The effect of the onsite repulsive U term in the DFT+U framework is to push the corresponding states in conduction and valence bands up and down, respectively. Hence, larger U parameter would push Cr 3$d$ states deep in the valence bands and even deeper than the Cu 3$d$ states, as found in Ref. 74. We believe the U parameter of 2 eV used in Ref. 76 is too small for Cu and Cr 3$d$ states, thus resulting in wrong band structures. In addition to DFT+U, hybrid functionals can also be employed to correct the band orders predicted by normal DFT. sX[35] and HSE06[76] calculations agree with the DFT+U (U = 4 eV for Cr) calculations with Cu 3$d$ states dominate the VBM. Our HSE (HSE03) calculations also reproduce the correct bands orders, as shown in Fig. 6b. We also check the band structure obtained from QS$GW$ calculations which should give the most accurate wave functions[70], and find it is consistent with that of HSE calculations qualitatively despite the large band gap of 4.5 eV at L. With the correct band order, DFT+U still underestimates the direct band gap as ~2.6 eV[74–76], while HSE06 overestimate it to 3.75 eV[76], compared to the experimental value of ~3.1-3.2 eV[6,68,69,77–79]. Although sX[35] and our HSE calculations predict reasonable band gaps of 3.1 eV and 3.2 eV, experiments[20] show that the exciton binding energy of CuCrO$_2$ is large as 0.33 eV.

To improve the description of the band structure of CuCrO$_2$, we adopt the $GW$ calculations based on the generalized Kohn-Sham (gKS) scheme[42]. The gKS+ $G_0W_0$ approach shows good performance for typical semiconductors or even compounds with shallow $d$ states. We use HSE wave functions as the starting point for the perturbative $GW_0$ calculations, since HSE reproduces the correct band orders. The $GW_0$@HSE band structure of CuCrO$_2$ is shown in Fig. 6c. We note here, that the $GW_0$ improvement of the band structure over $G_0W_0$ is quite small for the HSE starting point, only 0.04 eV increase of the band gap. The VBM is located at B site along LZ line for the up spin state, with 1 meV higher than A site. The fundamental direct band gap of 3.42 eV, which is also the lowest spin-allowed optical transition, is found at L for down spin state. The indirect band gap is 3.29 eV from B (up) to L (down). The dielectric function and oscillator strength are shown in Fig. 6d. The first absorption peak corresponds to a bound exciton with binding energy of 0.31 eV, in agreement with the experimental value of 0.33 eV[20]. The amplitude, as shown in Fig. 6c, indicates that the lowest excitation is mainly composed of Cu 3$d$-Cr 3$d$ transitions with a small contribution from Cu 3$d$-Cu 3$p$ transitions. However, this strong absorption peak is usually wrongly assigned to Cu 3$d$+O 2$p$ → Cu 3$d$+4$s$ transitions for delafossites in literatures due to incorrect band structures used.[15,17,20,78] We note that the Cu 3$d$+4$s$ states in conduction bands is located at Γ point, which is optical inactive due to dipole forbidden. According to our $GW$-BSE calculations, the lowest excitations are always at L point, mainly derived from either Cu 3$d$ → Cu 3$p$ (Al, Ga, In) or Cu 3$d$ → M 3$d$ (M = Sc, Y, Cr) transitions. The O 2$p$ states hybridize with Cu 3$d$ states in valence bands but with less contribution, we thus neglect the O 2$p$ in the notations.

FIG. 6 (a) PBE (b) HSE (c) $GW_0$@HSE band structures of CuCrO$_2$. The red circles and blue squares in (a) and (b) denote projections on Cr and Cu 3$d$ orbitals, respectively. The solid and dash lines in (c) denote the up and down spin states. The magenta circles show the fat-bands of the exciton. (d) dielectric function $\varepsilon_2$ and oscillator strength $\lambda$.

## IV. CONCLUSIONS

In conclusion, we calculate the band structures and the excitonic properties of delafossites CuMO$_2$ (M = Al, Ga, In, Sc, Y and Cr) by using the state-of-the-art $GW$-BSE approach. The HSE wave functions are used as the starting point of the $GW$ calculations for CuCrO$_2$, since HSE corrects the band orders wrongly predicted by PBE. For other delafossites, PBE wave functions are employed. We evaluate different self-consistency levels of $GW$ calculations on CuAlO$_2$, namely $G_0W_0$, $GW_0$, $GW$ and QS$GW$, and find that the $GW_0$ scheme with self-consistent updates on the Green's function while keeping the screened Coulomb potential on RPA level yields band gaps closest to experiments. Our results show that all of the delafossites are indirect band gap semiconductors with large exciton binding energies, varying from 0.24 to 0.44 eV. The lowest and strongest exciton absorption is always located at L point of the rhombohedral Brillouin zone, mainly contributed from either Cu 3$d$ → Cu3$p$ (Al, Ga, In) or Cu 3$d$ → M 3$d$ (M = Sc, Y, Cr) transitions. Taking the electron-hole effect into account, the calculated band gaps agree quite well with experiments.

## ACKNOWLEDGEMENTS

This work was funded in part by the Office of Energy Efficiency and Renewable Energy (EERE), U.S. Department of Energy, under Award Number DE-EE0006712, the Ohio Research Scholar Program, and the National Science Foundation under contract no. CHE−1230246 and DMR−1534686. This research used the resources of the National Energy Research Scientific Computing Center, which is supported by the Office of Science of the U.S. Department of Energy under Contract No. DE-AC02-05CH11231.

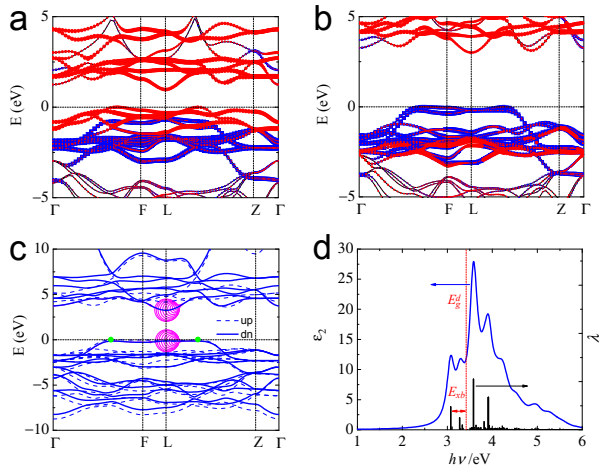